\documentclass{ws-rv9x6}
\usepackage{subfigure}   
\usepackage{ws-rv-thm}   
\usepackage{ws-rv-van}   
\makeindex

\begin{document}

\chapter[Partial restoration of chiral symmetry in hot and dense neutron matter]{Partial restoration of chiral symmetry in \\hot and dense neutron matter}\label{ra_ch1}

\author[J. W. Holt]{Jeremy W. Holt\footnote{Corresponding author: holt@physics.tamu.edu}}

\address{Cyclotron Institute and Dept.\ of Physics and Astronomy\\
Texas A\&M University, College Station, TX}

\author[J. W. Holt and K. E. Rockcliffe]{Keighley E. Rockcliffe}

\address{Department of Physics, Applied Physics, and Astronomy\\
Rensselaer Polytechnic Institute, Troy, NY}

\begin{abstract}
We review efforts to describe the approach to chiral symmetry restoration
in neutron matter from the low-energy realization of QCD, chiral
effective field theory.
\end{abstract}
\body


\section{Introduction}

The restoration of chiral symmetry in hot and dense hadronic matter and the associated 
observable signatures in relativistic heavy ion collisions were favorite research interests of 
Gerry Brown for two decades\cite{brown88,brown91,brown02,brown07,brown09}. 
Gerry attacked the problem with his characteristic directness, making an 
early prediction \cite{brown88,brown91} with Mannque Rho that nucleons and heavy 
mesons shed mass in the approach to chiral symmetry restoration according to a 
universal scaling law. At the time that one of the authors (JWH) became Gerry's graduate 
student at Stony Brook in the mid-2000's, numerous experimental investigations 
of the Brown-Rho scaling conjecture were underway \cite{trnka05,nasseripour07,kotulla08}.
Despite the lack of clean experimental signatures\cite{brown07,brown09} for Brown-Rho 
scaling at finite density and temperature, understanding the effects of medium-modified 
meson masses and couplings on the nucleon-nucleon 
potential was nevertheless an inspiring topic for a PhD thesis. 

One of Gerry's favorite 
words of wisdom for young graduate students was, ``Build simple models and don't 
let anyone tell you it has to be more complicated.'' The present contribution to this memorial
volume will focus on a somewhat ``more complicated'' method to understand medium-modified 
nucleon-nucleon (NN) interactions and the approach to chiral symmetry restoration in hot and 
dense neutron matter. This framework, utilizing microscopic chiral effective field theory (EFT) two- and 
three-body nuclear forces, in fact produces in-medium two-body interactions
qualitatively similar to models incorporating Brown-Rho scaling
\cite{friman96,holt07,holt08,holt09,holt11npa,holt12,dong13}. When employed in many-body
perturbation theory to calculate the neutron matter thermodynamic equation of state, the 
coarse-resolution chiral potentials used in the present work show excellent convergence 
properties. Here again we are indebted to the work of Gerry Brown, who together with Tom Kuo, 
Scott Bogner and Achim Schwenk, pioneered the use of low-momentum nucleon-nucleon 
potentials \cite{bogner03} derived from effective interaction theory and the renormalization group.

Chiral symmetry generically appears in quantum field theories with massless fermions. 
In the chiral limit of QCD the left- and right-handed quarks decouple and the Lagrangian is 
invariant under independent $SU(2)_{L,R}$ transformations in flavor space. In addition to
the explicit breaking of chiral symmetry due to the small but nonzero bare quark mass
arising from coupling to the Higgs field, the strong attraction between quark-antiquark pairs 
leads to the formation of a scalar quark condensate $\langle 0 | \bar{q} q | 0 \rangle$ 
in the QCD vacuum that spontaneously breaks
chiral symmetry. At the high temperatures and/or densities encountered in core-collapse 
supernovae or neutron star mergers, however, chiral symmetry may be restored \cite{alford05,fischer11}.

At low densities the quark condensate in nuclear matter decreases linearly \cite{drukarev90,cohen92}
and proportional to the pion-nucleon sigma term $\sigma_{\pi N} = m_q \partial M_N / \partial m_q$, 
which encodes the small change in the nucleon mass $M_N$ from the explicit breaking
of chiral symmetry. Neglecting interaction contributions to the ground state energy, the linear term 
alone gives rise to chiral symmetry restoration in 
cold symmetric nuclear matter and pure neutron matter at a density $n \simeq 2.5 n_0$, 
assuming a value of $\sigma_{\pi N} = 45$\,MeV\cite{gasser91} (note that a more recent chiral EFT 
analysis\cite{alarcon12} has found the larger value of $\sigma_{\pi N} = 59 \pm 7$\,MeV), 
where $n_0 = 0.16$\,fm$^{-3}$.
This is in the vicinity of the energy density at which chiral symmetry is restored at finite temperature
and zero net baryon density from lattice QCD \cite{hegde14}.
Corrections to the leading density dependence have been obtained from the quark-mass dependence
of interaction contributions to the nuclear matter ground state energy in one-boson-exchange 
models \cite{li94,brockmann96}, 
in-medium chiral perturbation theory\cite{kaiser08,kaiser09}, and by employing high-precision 
chiral nuclear forces at next-to-next-to-next-to-leading order (N3LO)\cite{krueger13}. In all 
of these studies, nuclear mean fields and correlations arising from two- and three-body interactions 
consistently suppress chiral symmetry restoration beyond nuclear matter saturation density, 
relative to the noninteracting case.

At finite temperature, chiral perturbation theory was used to predict a phase transition to 
chiral symmetry restored matter at a critical temperature of $T_c \simeq 190$\,MeV\cite{gerber89}, which is 
somewhat larger than the presently accepted value of $T_c = 155 \pm 10$\,MeV from lattice 
QCD\cite{borsanyi10,bazavov12,bhattacharya14}. In the present study we compute both the 
temperature and density dependence of the scalar quark condensate in neutron matter from
high-precision two- and three-nucleon interactions. Previous finite-temperature calculations of
the quark condensate in symmetric nuclear matter in the framework of in-medium chiral perturbation 
theory \cite{fiorilla12b} revealed that thermal effects wash out the interaction contributions that tend
to delay chiral symmetry restoration at high densities. This leads to a nearly 
linear density dependence of the condensate for temperatures greater
than $T = 50$\,MeV. Hot proto-neutron stars born immediately after core-collapse supernovae or 
the hypermassive neutron stars that exist transiently after the merger of two neutron 
stars may therefore be more compelling candidate sites for quark-hadron phase transitions
than cold neutron stars.

\section{Neutron matter at finite temperature from many-body perturbation theory}

Chiral effective field theory is the appropriate tool to study hadronic matter at the scales relevant 
in nuclear astrophysics (well below the chiral symmetry breaking scale of 
$\Lambda_\chi \simeq 1$\,GeV). We start from a coarse-resolution chiral 
potential\cite{coraggio13,coraggio14} with a 
momentum-space cutoff of $\Lambda = 414$\,MeV, which has been shown to exhibit good 
convergence properties \cite{coraggio13,coraggio14,wellenhofer14,wellenhofer15,wellenhofer16}
in many-body perturbation theory calculations of infinite nuclear matter, comparable to 
low-momentum potentials constructed via renormalization group methods 
\cite{bogner03,bogner05,hebeler10,hebeler11}. In the present calculation the free energy
per particle of pure neutron matter at finite temperature is computed in the 
imaginary-time Matsubara formalism. The perturbation series for the grand canonical
potential $\Omega$ reads
\begin{equation}
\Omega(\mu,T) = \Omega_0(\mu,T) + \lambda \Omega_1(\mu,T) + \lambda^2 \Omega_2(\mu,T)
+ \cdots,
\label{pto}
\end{equation}
where $\lambda$ is an arbitrary strength parameter, $T$ is the temperature, and $\mu$ is the
chemical potential. Eq.\ (\ref{pto}) 
can be reformulated in the canonical ensemble through the Kohn-Luttinger-Ward
prescription \cite{kohn60,luttinger60}. The result is a rearrangement of the perturbation series:
\begin{equation}
F(\mu_0,T) = F_0(\mu_0,T) + \lambda \Omega_1(\mu_0,T) + \lambda^2 \left ( \Omega_2(\mu_0,T)
- \frac{1}{2} \frac{(\partial \Omega_1 / \partial \mu_0)^2}{\partial^2 \Omega_0 / \partial \mu_0^2 } \right ) + \cdots,
\label{klw}
\end{equation}
where $\mu_0$ is the chemical potential of the noninteracting
system. There is a one-to-one correspondence between the nucleon density $n$ and
the effective chemical potential $\mu_0$ through 
\begin{equation}
n(\mu_0,T) = -\frac{\partial \Omega_0}{\partial \mu_0},
\end{equation}
where the noninteracting grand canonical potential has the well known form
\begin{equation}
\Omega_0(\mu_0,T) = -\frac{1}{\beta}\frac{1}{\pi^2}\int_0^\infty dp \, p^2 \log (1+e^{-\beta(e(p)-\mu_0)})
\end{equation}
and $e(p) = p^2/2M_N$ is the single-particle energy.
The first term in the expansion of the free energy is related to $\Omega_0$ via
$F_0(\mu_0,T) = \Omega_0(\mu_0,T) + n \mu_0$.

\begin{figure}[t]
\centerline{\includegraphics[width=6.5cm]{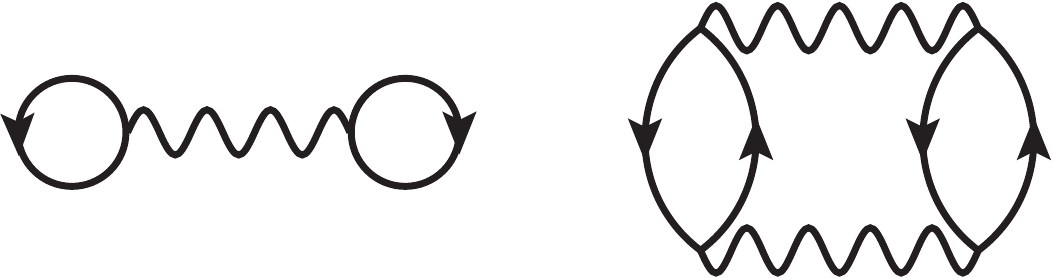}}
\caption{First- and second-order diagrammatic contributions to the free energy of pure
neutron matter from n3lo414 chiral two- and three-nucleon forces. The wavy line represents
a density- and temperature-dependent NN interaction derived from the chiral three-body force.} 
\label{gold}
\end{figure}

The first- and second-order perturbative contributions to the free energy of neutron
matter are shown diagrammatically in Fig.\ \ref{gold}. The wavy lines represent the sum of the free-space
NN interaction and an in-medium NN potential 
constructed from the N2LO chiral three-body force by summing one leg over the filled Fermi sea of
noninteracting neutrons. This approximation can be improved upon by including three-body forces
at N3LO \cite{tews13} and by treating three-body forces explicitly at higher orders in perturbation theory\cite{kaiser12,drischler16}. In the present case, the effective interaction depends on the density and 
temperature of neutron matter. The explicit expressions for the diagrams shown in Fig.\
\ref{gold} are given by
\begin{equation}
\Omega_{1}(\mu_0,T) = \frac{1}{2}\sum_{\sigma_1,\sigma_2}  
\!\! \int  \! \! \frac{d^3k_1 d^3k_2}{(2\pi)^6}\, \,
n_{ k_1} n_{k_2} \langle 1 2 \left | (\bar{V}_{NN}+ \bar{V}_{NN}^{med}/3)
\right | 1 2 \rangle,
\label{diagramexp1NN}
\end{equation}
\begin{eqnarray}
\Omega_2^n(\mu_0,T) &=& -\frac{1}{8} 
\prod_{i=1}^4 \left( \sum_{\sigma_i}
\int  \! \! \frac{d^3k_i }{\left(2 \pi \right)^3} \right) \left(2 \pi \right)^3
\delta \left(  \vec{k}_1 +\vec{k}_2-\vec{k}_3-\vec{k}_4 \right) \\ \nonumber  
&\times& \frac{n_{k_1}n_{k_2} \bar{n}_{k_3} \bar{n}_{k_4}-
\bar{n}_{k_1} \bar{n}_{k_2}n_{k_3}n_{k_4}}
{e_3+e_4-e_1-e_2} 
\left| \langle 1 2 \left | (\bar{V}_{NN}+\bar{V}_{NN}^{med}) \right | 3 4 \rangle \right |^2,
\label{diagramexp2n}
\end{eqnarray}
where $n_k$ is the Fermi distribution function, $\bar n_k=1-n_k$, and $\bar{V}$ is the 
antisymmetrized potential. We have written only the expression for the ``normal'' second-order
contribution $\Omega_2^n$ and neglected the anomalous contribution, which was 
shown\cite{wellenhofer14} to have a negligible effect when combined with the derivative
term in Eq.\ (\ref{klw}). The free energy of neutron matter has been computed 
previously\cite{wellenhofer15} employing the n3lo414 chiral potential. 
At low densities and high temperatures, the results were 
found to agree well with the model-independent virial expansion, and in the vicinity of
nuclear matter saturation density the symmetry energy 
$E_{sym} = 32.5$\,MeV and slope $L = 53.8$\,MeV are consistent with empirical constraints 
\cite{dutra12}. 

From the Hellmann-Feynman theorem\cite{cohen92,fiorilla12b}
\begin{equation}
\langle \psi(\alpha) | \frac{d}{d\alpha} H(\alpha) | \psi(\alpha \rangle
= \frac{d}{d\alpha}E(\alpha),
\end{equation}
the chiral condensate in finite-density matter is related to the vacuum value by choosing 
$\alpha = m_q$:
\begin{equation}
\langle \psi | \bar q q | \psi \rangle - \langle 0 | \bar q q | 0 \rangle = 
\frac{1}{2} \frac{d}{dm_q} \left [ M_N n + {\cal E} \right ],
\end{equation}
where $| \psi \rangle$ is the neutron matter ground state and ${\cal E}$ is the energy density.
Since we do not have direct access to the quark-mass dependence of the neutron matter
ground state energy, we employ the Gell-mann--Oakes--Renner relation at leading order
\begin{equation}
m_\pi^2 f_\pi^2 \simeq -2 \bar{m_q} \langle 0 | \bar{q} q | 0 \rangle,
\end{equation}
where $\bar{m}_q$ is the average light-quark mass and $f_\pi = 92.4$\,MeV is the pion decay 
constant, to write the chiral condensate in terms of the pion-mass dependence of the energy
per particle $\bar E$:
\begin{equation}
\frac{\langle \bar q q \rangle_n}{\langle \bar q q \rangle_0} = 1 - \frac{n}{f_\pi^2} 
\left [ \frac{\sigma_{\pi N}} {m_\pi^2} + \frac{d \bar E}{dm_\pi^2}  \right ].
\label{cc0t}
\end{equation}
This expression can be extended to finite temperature\cite{fiorilla12b} by replacing the 
energy per particle in Eq.\ (\ref{cc0t}) with the free energy per particle $\bar F$
\begin{equation}
\frac{\langle \bar{q} q \rangle_{n,T}}{\langle \bar{q} q \rangle_{0,0}} = 
1 - \frac{n}{f_\pi^2}\frac{\partial \bar F(n,T)}{\partial m_\pi^2},
\end{equation}
and absorbing the nucleon mass and chemical potential $\mu_0$ into $\bar F$.

\section{Results}

\begin{figure}[t]
\centerline{\includegraphics[width=8cm,angle=270]{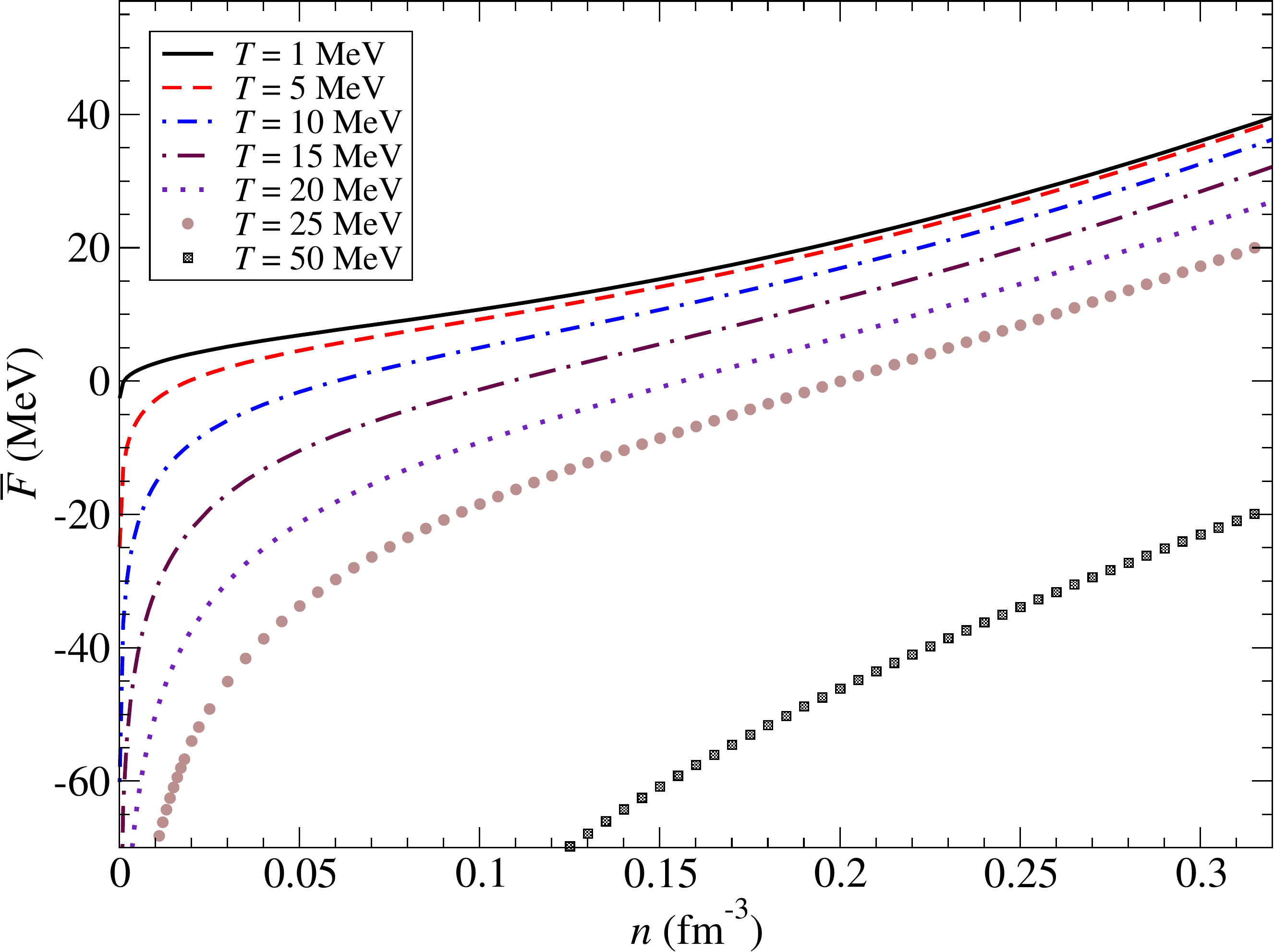}}
\caption{Free energy per particle $\bar F$ of pure neutron matter as a function of density for different isotherms.
Results are obtained at second order in many-body perturbation theory employing the coarse-resolution 
chiral potential n3lo414 described in the text.} 
\label{fpnm}
\end{figure}

We begin by plotting in Fig.\ \ref{fpnm} the neutron matter thermodynamic equation of state (the 
free energy per particle as a function of temperature and density) obtained from the 
n3lo414 chiral two- and three-body potentials. The results are in good agreement with 
previous studies \cite{wellenhofer15,wellenhofer16}, which employed the same two- and
three-body potentials but included also self-consistent Hartree-Fock single-particle energies
in the second-order diagrams. As we will show later, second-order perturbative contributions
to the neutron matter equation of state and the chiral condensate are relatively small. 
The inclusion of in-medium nucleon propagators will therefore be postponed to future work
and higher-order contributions to the nucleon single-particle potential will be 
addressed.

\begin{figure}[t]
\centerline{\includegraphics[width=10cm]{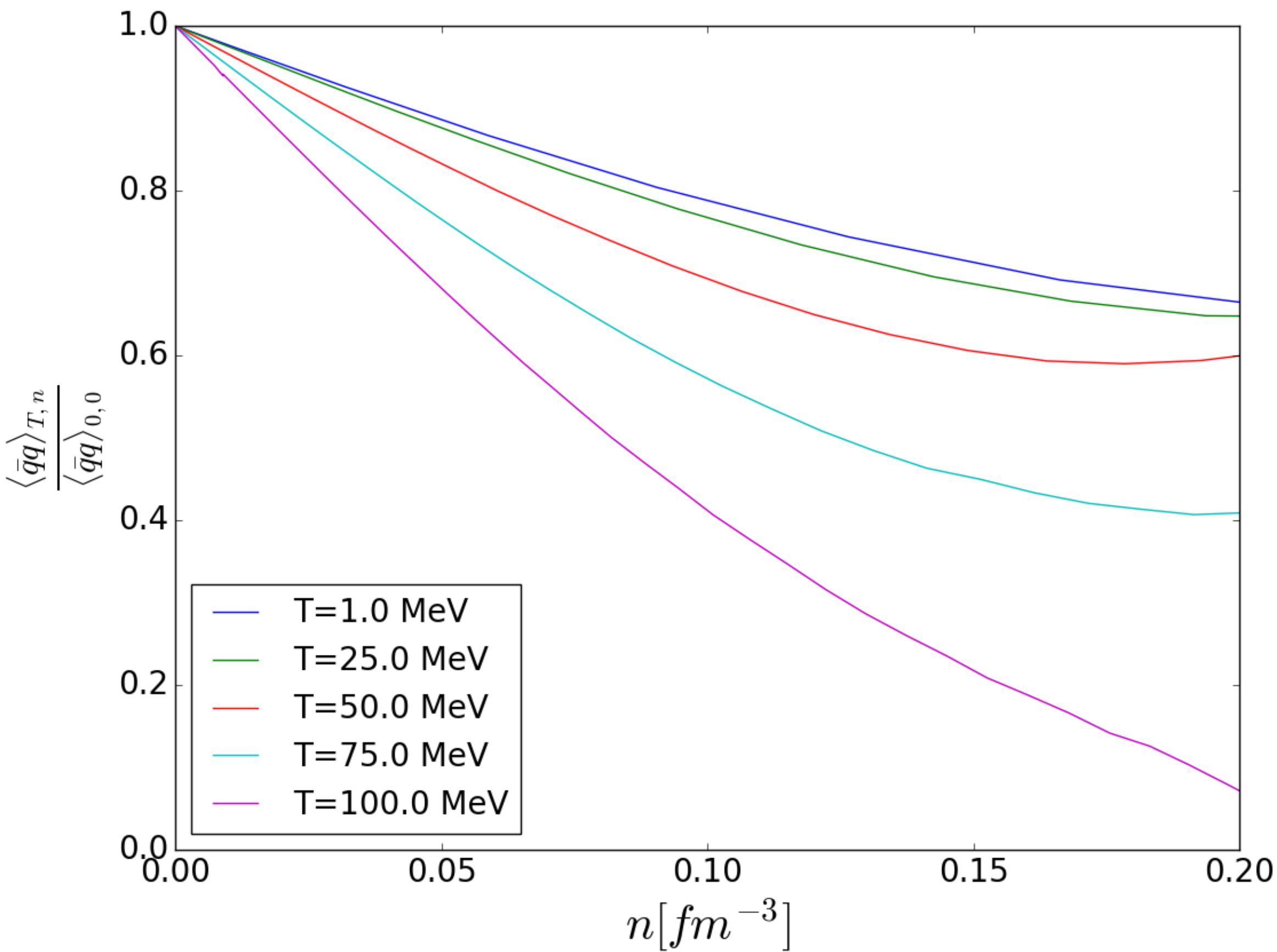}}
\caption{Ratio of the chiral condensate in pure neutron matter to that in the vacuum 
as a function of density and temperature from the pion-mass dependence of the free energy 
computed in many-body perturbation theory with coarse-resolution chiral two- and three-body
potentials.} 
\label{cc}
\end{figure}

In Fig.\ \ref{cc} we show the ratio of the chiral condensate in neutron matter at finite temperature $T$
to the chiral condensate in vacuum as a function of the nucleon number density $n$. In computing the 
numerical derivatives of the different contributions to the free energy density with respect to the
pion mass, we chose to vary the pion mass by 5\%, 2\%, and 1\%. We found only very small
differences in the values of the numerical derivatives, and in the figure the results from a 1\% 
change in the pion mass are shown. We observe that at zero temperature the density dependence 
of the chiral condensate falls within the uncertainty band obtained in Ref.\ \cite{krueger13}, which 
included as well three-body forces at N3LO in the chiral power counting. 

Temperatures up to 100\, MeV are considered in the present work. In the highest temperature
regime, the presence of thermal pions enhance the trend toward chiral symmetry restoration
\cite{gerber89,toublan97,fiorilla12b}, leading to a qualitative difference in the condensate ratio 
especially at low density. In the present work, we focus on nuclear interaction contributions to 
the medium dependence of the chiral condensate and will include effects due to thermal pions 
in future work. We neglect also the quark-mass dependence of the axial coupling strength 
$g_A$, pion decay constant $f_\pi$, and short-range low-energy constants in the 2N and 3N 
sectors. For zero-temperature neutron matter, these effects were estimated\cite{krueger13} to 
contribute on the order of 25\% to the theoretical uncertainty in the chiral condensate.
More detailed discussions can be found in the literature\cite{bernard06,berengut13}. 

As found in previous chiral effective field theory calculations \cite{kaiser09,krueger13} 
of the pion-mass dependence of the neutron matter ground state energy at zero temperature,
we observe that interaction
contributions result in a relatively small change in the leading linear decrease of the chiral
condensate with increasing density. In particular, repulsive interactions 
delay the approach to chiral symmetry restoration and increase with nucleon density. 
At $n=0.2$\,fm$^{-3}$ nuclear interactions increase the chiral condensate by about 15\%.
This is a larger effect, though qualitatively similar, to what has been observed employing
one-boson-exchange models of the nucleon-nucleon interaction\cite{li94}.

\begin{figure}[t]
\centerline{\includegraphics[width=10cm]{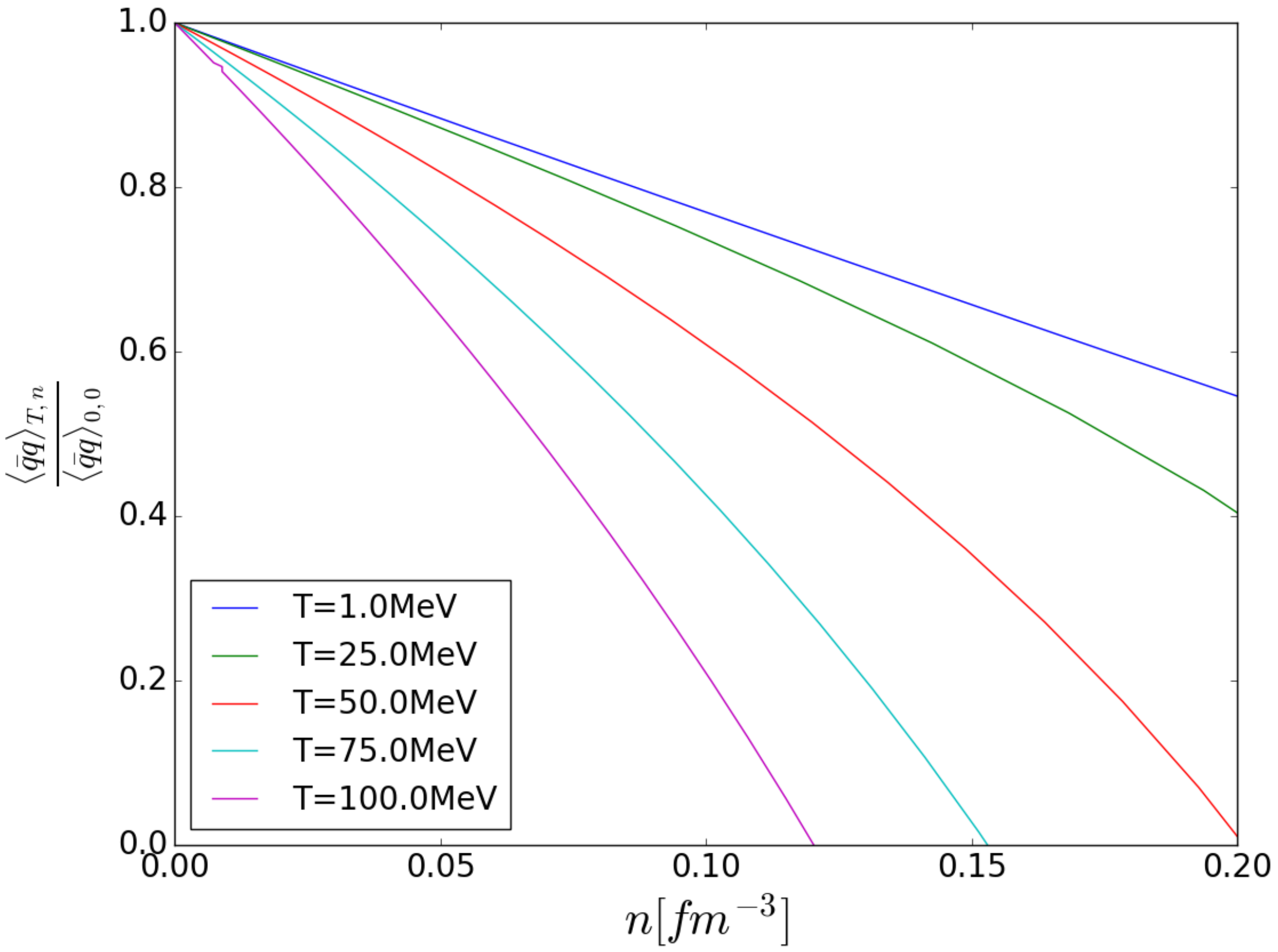}}
\caption{Same as in Fig.\ \ref{cc}, except that only the noninteracting contribution to the 
medium dependence of the chiral condensate is included.}
\label{cc0}
\end{figure}

\begin{figure}[t]
\centerline{\includegraphics[width=10cm]{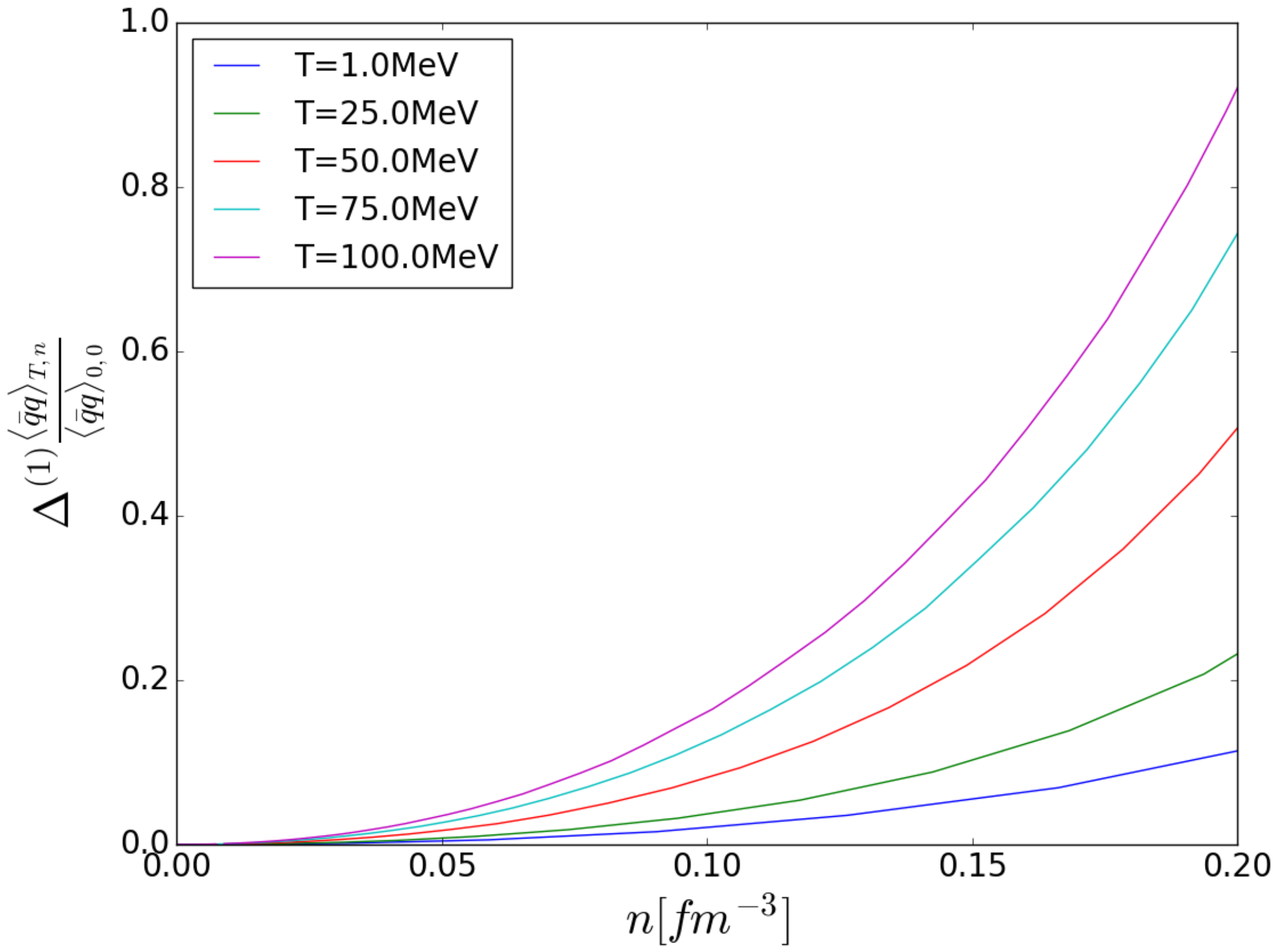}}
\caption{Change in the chiral condensate ratio as a function of temperature and density including
the first-order perturbative contribution to the free energy of neutron matter.} 
\label{cc1}
\end{figure}

\begin{figure}[t]
\centerline{\includegraphics[width=10cm]{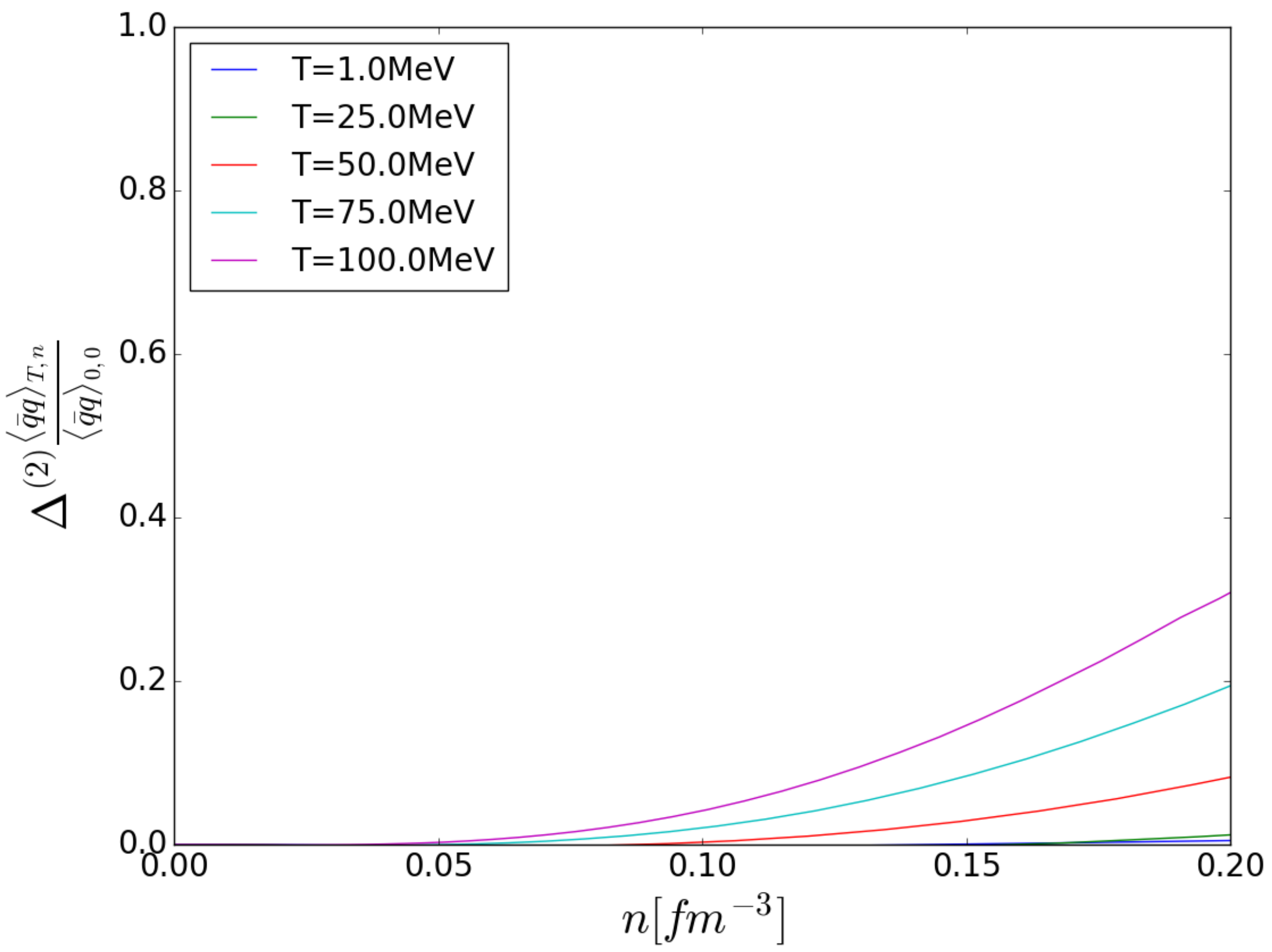}}
\caption{Same as in Fig.\ \ref{cc1}, except that only the second-order perturbative contribution to the
free energy has been included.} 
\label{cc2}
\end{figure}

Increasing the 
temperature, on the other hand, is highly effective at promoting chiral symmetry restoration
in neutron matter.
This comes predominantly from the noninteracting contributions to the pion mass
dependence of the neutron matter free energy, as shown in Fig.\ \ref{cc0}.
Increasing the quark mass enhances both the value of the nucleon mass and the kinetic 
energy contribution to the neutron matter free energy. 
In Figs.\ \ref{cc1} and \ref{cc2} we plot the first-order, $\Delta^{(1)}\frac{\langle 
\bar q q \rangle_{T,n}}{\langle \bar{q} q \rangle_{0,0}}$, and second-order, 
$\Delta^{(2)}\frac{\langle \bar q q \rangle_{T,n}}{\langle \bar{q} q \rangle_{0,0}}$,
perturbative contributions to the chiral condensate ratio from two- and three-nucleon 
chiral forces. For all densities and temperatures considered in the present work, both 
the first- and second-order contributions tend to delay the restoration of chiral symmetry
for increasing density and temperature. This feature was observed already in Ref.\ \cite{krueger13} for
neutron matter at zero temperature, and we note that it holds with increasing temperature.
The leading Hartree-Fock contribution is consistently about
three times larger in magnitude than the second-order correction. For temperatures up to 
$T=50$\,MeV and densities below $n=0.2$\,fm$^{-3}$, the second-order interaction contribution 
to the scalar quark condensate gives less than a 10\% correction. This is in agreement 
with previous studies of the neutron matter equation of state where the use of coarse-resolution 
chiral potentials strongly reduced the effect of the second-order bubble diagram. Uncertainties 
associated with the pion-nucleon sigma term and low-energy constants in
two- and three-nucleon forces are therefore expected to dominate the theoretical errors\cite{krueger13}.

\section{Summary}

Many-body perturbation theory with coarse-resolution chiral potentials now allows for the 
systematic study of infinite nuclear matter properties with reduced theoretical uncertainties.
In the present work we have calculated the scalar quark condensate in neutron matter at finite
temperature employing a recently developed chiral nuclear potential with a momentum-space cutoff
of $\Lambda = 414$\,MeV. We considered separately the pion-mass dependence of noninteracting 
contributions to the free energy of neutron matter as well as interaction effects from two- and
three-body nuclear forces. For all densities and temperatures, the noninteracting contributions are
dominant and lead to a decrease of the chiral condensate with increasing density and temperature.
In the absence of nuclear interactions, hot neutron matter around nuclear matter saturation density
should already exhibit a transition to a chiral symmetry restored phase. However, interaction effects
generically delay chiral symmetry restoration and increase in magnitude with both the density
and temperature.

\section{Acknowledgments}
K. E. Rockcliffe acknowledges the kind hospitality of the Texas A\&M Cyclotron Institute. 
Work supported in part by NSF grant PHY-1263281 and DOE grant DE-FG03-93ER40773.

\bibliographystyle{ws-rv-van}

\bibliography{biblio}

\begin{thebibliography}{50}
\providecommand{\natexlab}[1]{#1}
\providecommand{\url}[1]{\texttt{#1}}
\expandafter\ifx\csname urlstyle\endcsname\relax
  \providecommand{\doi}[1]{doi: #1}\else
  \providecommand{\doi}{doi: \begingroup \urlstyle{rm}\Url}\fi

\bibitem{brown88}
G.~E. Brown, \emph{Nucl. Phys.} {\bf A488}, \penalty0 689  (1988).

\bibitem{brown91}
G.~E. Brown and M.~Rho, \emph{Phys. Rev. Lett.} {\bf 66}, \penalty0 2720
  (1991).

\bibitem{brown02}
G.~E. Brown and M.~Rho, \emph{Phys. Rept.} {\bf 363}, \penalty0 85  (2002).

\bibitem{brown07}
G.~E. Brown, J.~W. Holt, and M.~Rho, \emph{Phys. Rept.} {\bf 439}, \penalty0
  161  (2007).

\bibitem{brown09}
G.~E. Brown, M.~Harada, J.~W. Holt, M.~Rho, and C.~Sasaki, \emph{Prog. Theor.
  Phys.} {\bf 121}, \penalty0 1209  (2009).

\bibitem{trnka05}
D.~Trnka et~al., \emph{Phys. Rev. Lett.} {\bf 94}, \penalty0 192303  (2005).

\bibitem{nasseripour07}
R.~Nasseripour et~al., \emph{Phys. Rev. Lett.} {\bf 99}, \penalty0 262302
  (2007).

\bibitem{kotulla08}
M.~Kotulla et~al., \emph{Phys. Rev. Lett.} {\bf 100}, \penalty0 192302  (2008).

\bibitem{friman96}
B.~Friman and M.~Rho, \emph{Nucl. Phys.} {\bf A606}, \penalty0 303  (1996).

\bibitem{holt07}
J.~W. Holt, G.~E. Brown, J.~D. Holt, and T.~T.~S. Kuo, \emph{Nucl. Phys.} {\bf
  A785}, \penalty0 322  (2007).

\bibitem{holt08}
J.~W. Holt, G.~E. Brown, T.~T.~S. Kuo, J.~D. Holt, and R.~Machleidt,
  \emph{Phys. Rev. Lett.} {\bf 100}, \penalty0 062501  (2008).

\bibitem{holt09}
J.~W. Holt, N.~Kaiser, and W.~Weise, \emph{Phys. Rev. C}. {\bf 79}, \penalty0
  054331  (2009).

\bibitem{holt11npa}
J.~W. Holt, N.~Kaiser, and W.~Weise, \emph{Nucl. Phys.} {\bf A870-871},
  \penalty0 1  (2011).

\bibitem{holt12}
J.~W. Holt, N.~Kaiser, and W.~Weise, \emph{Nucl. Phys.} {\bf A876}, \penalty0
  61  (2012).

\bibitem{dong13}
H.~Dong, T.~T.~S. Kuo, H.~K. Lee, R.~Machleidt, and M.~Rho, \emph{Phys. Rev.
  C}. {\bf 87}, \penalty0 054332  (2013).

\bibitem{bogner03}
S.~K. Bogner, T.~T.~S. Kuo, and A.~Schwenk, \emph{Phys. Rept.} {\bf 386},
  \penalty0 1  (2003).

\bibitem{alford05}
M.~Alford, M.~Braby, M.~Paris, and S.~Reddy, \emph{Astrophys. J.} {\bf 629},
  \penalty0 969  (2005).

\bibitem{fischer11}
T.~Fischer, I.~Sagert, G.~Pagliara, M.~Hempel, J.~Schaffner-Bielich,
  T.~Rauscher, F.-K. Thielemann, R.~K\"{a}ppeli, G.~Mart\'{i}nez-Pinedo, and
  M.~Liebend\"{o}rfer, \emph{Astrophys. J. Suppl.} {\bf 194}, \penalty0 39
  (2011).

\bibitem{drukarev90}
E.~G. Drukarev and E.~Levin, \emph{Nucl. Phys.} {\bf A511}, \penalty0 679
  (1990).

\bibitem{cohen92}
T.~D. Cohen, R.~J. Furnstahl, and D.~G. Griegel, \emph{Phys. Rev. C}. {\bf 45},
  \penalty0 1881  (1992).

\bibitem{gasser91}
J.~Gasser, H.~Leutwyler, and M.~E. Sainio, \emph{Phys. Lett.} {\bf B253},
  \penalty0 252  (1991).

\bibitem{alarcon12}
J.~M. Alarc\'on, J.~M. Camalich, and J.~A. Oller, \emph{Phys. Rev. D}. {\bf
  85}, \penalty0 051503  (2012).

\bibitem{hegde14}
P.~Hegde and (BNL-Bielefeld-CCNU), \emph{Proceedings, 32nd International
  Symposium on Lattice Field Theory (Lattice 2014)}. {\bf PoS LATTICE2014},
  \penalty0 226  (2014).

\bibitem{li94}
G.~Q. Li and C.~M. Ko, \emph{Phys. Lett. B}. {\bf 338}, \penalty0 118  (1994).

\bibitem{brockmann96}
R.~Brockmann and W.~Weise, \emph{Phys. Lett. B}. {\bf 367}, \penalty0 40
  (1996).

\bibitem{kaiser08}
N.~Kaiser, P.~de~Homont, and W.~Weise, \emph{Phys. Rev. C}. {\bf 77}, \penalty0
  025204  (2008).

\bibitem{kaiser09}
N.~Kaiser and W.~Weise, \emph{Phys. Lett. B}. p. 671  (2009).

\bibitem{krueger13}
T.~Kr\"{u}ger, I.~Tews, B.~Friman, K.~Hebeler, and A.~Schwenk, \emph{Phys.
  Lett. B}. {\bf 726}, \penalty0 412  (2013).

\bibitem{gerber89}
P.~Gerber and H.~Leutwyler, \emph{Nucl. Phys.} {\bf B321}, \penalty0 387
  (1989).

\bibitem{borsanyi10}
S.~Bors{\'a}nyi, Z.~Fodor, C.~Hoelbling, S.~D. Katz, S.~Krieg, C.~Ratti, and
  K.~K. Szab{\'o}, \emph{Journal of High Energy Physics}. {\bf 2010}\penalty0
  (9), \penalty0 1  (2010).

\bibitem{bazavov12}
A.~Bazavov, T.~Bhattacharya, M.~Cheng, C.~DeTar, H.-T. Ding, S.~Gottlieb,
  R.~Gupta, P.~Hegde, U.~M. Heller, F.~Karsch, E.~Laermann, L.~Levkova,
  S.~Mukherjee, P.~Petreczky, C.~Schmidt, R.~A. Soltz, W.~Soeldner, R.~Sugar,
  D.~Toussaint, W.~Unger, and P.~Vranas, \emph{Phys. Rev. D}. {\bf 85},
  \penalty0 054503  (2012).

\bibitem{bhattacharya14}
T.~Bhattacharya, M.~I. Buchoff, N.~H. Christ, H.-T. Ding, R.~Gupta, C.~Jung,
  F.~Karsch, Z.~Lin, R.~D. Mawhinney, G.~McGlynn, S.~Mukherjee, D.~Murphy,
  P.~Petreczky, D.~Renfrew, C.~Schroeder, R.~A. Soltz, P.~M. Vranas, and
  H.~Yin, \emph{Phys. Rev. Lett.} {\bf 113}, \penalty0 082001  (2014).

\bibitem{fiorilla12b}
S.~Fiorilla, N.~Kaiser, and W.~Weise, \emph{Phys. Lett. B}. {\bf 714},
  \penalty0 251  (2012).

\bibitem{coraggio13}
L.~Coraggio, J.~W. Holt, N.~Itaco, R.~Machleidt, and F.~Sammarruca, \emph{Phys.
  Rev. C}. {\bf 87}, \penalty0 014322  (2013).

\bibitem{coraggio14}
L.~Coraggio, J.~W. Holt, N.~Itaco, R.~Machleidt, L.~E. Marcucci, and
  F.~Sammarruca, \emph{Phys. Rev. C}. {\bf 89}, \penalty0 044321  (2014).

\bibitem{wellenhofer14}
C.~Wellenhofer, J.~W. Holt, N.~Kaiser, and W.~Weise, \emph{Phys. Rev. C}. {\bf
  89}, \penalty0 064009  (2014).

\bibitem{wellenhofer15}
C.~Wellenhofer, J.~W. Holt, and N.~Kaiser, \emph{Phys. Rev. C}. {\bf 92},
  \penalty0 015801  (2015).

\bibitem{wellenhofer16}
C.~Wellenhofer, J.~W. Holt, and N.~Kaiser, \emph{Phys. Rev. C}. {\bf 93},
  \penalty0 055802  (2016).

\bibitem{bogner05}
S.~K. Bogner, A.~Schwenk, R.~J. Furnstahl, and A.~Nogga, \emph{Nucl. Phys.}
  {\bf A763}, \penalty0 59  (2005).

\bibitem{hebeler10}
K.~Hebeler and A.~Schwenk, \emph{Phys. Rev. C}. {\bf 82}, \penalty0 014314
  (2010).

\bibitem{hebeler11}
K.~{Hebeler}, S.~K. {Bogner}, R.~J. {Furnstahl}, A.~{Nogga}, and A.~{Schwenk},
  \emph{Phys. Rev. C}. {\bf 83}, \penalty0 031301  (2011).

\bibitem{kohn60}
W.~Kohn and J.~M. Luttinger, \emph{Phys. Rev.} {\bf 118}, \penalty0 41  (1960).

\bibitem{luttinger60}
J.~M. Luttinger and J.~C. Ward, \emph{Phys. Rev.} {\bf 118}, \penalty0 1417
  (1960).

\bibitem{tews13}
I.~Tews, T.~Kr{\"u}ger, K.~Hebeler, and A.~Schwenk, \emph{Phys. Rev. Lett.}
  {\bf 110}, \penalty0 032504  (2013).

\bibitem{kaiser12}
N.~Kaiser, \emph{Eur. J. Phys. A}. {\bf 48}, \penalty0 58  (2012).

\bibitem{drischler16}
C.~Drischler, A.~Carbone, K.~Hebeler, and A.~Schwenk, \emph{arXiv:1608.05615}
  (2016).

\bibitem{dutra12}
M.~Dutra, O.~Louren\ifmmode~\mbox{\c{c}}\else \c{c}\fi{}o, J.~S. S\'a~Martins,
  A.~Delfino, J.~R. Stone, and P.~D. Stevenson, \emph{Phys. Rev. C}. {\bf 85},
  \penalty0 035201  (2012).

\bibitem{toublan97}
D.~Toublan, \emph{Phys. Rev. D}. {\bf 56}, \penalty0 5629  (1997).

\bibitem{bernard06}
V.~Bernard and U.-G. Meissner, \emph{Phys. Lett. B}. {\bf 639}, \penalty0 278
  (2006).

\bibitem{berengut13}
J.~C. Berengut, E.~Epelbaum, V.~V. Flambaum, C.~Hanhart, U.-G. Mei\ss{}ner,
  J.~Nebreda, and J.~R. Pel\'aez, \emph{Phys. Rev. D}. {\bf 87}, \penalty0
  085018  (2013).

\end{thebibliography}


\end{document}